# Tunable magnetic response of metamaterials

Shumin Xiao[1], Uday K. Chettiar[1], Alexander V. Kildishev[1], Vladimir Drachev[1], I. C. Khoo[2] and Vladimir M. Shalaev[1,a)]

[1]School of Electrical and Computer Engineering and Birck Nanotechnology Center, Purdue University, West Lafayette, Indiana 47907, USA

[2]Electrical Engineering Department, Pennsylvania State University, University Park, PA 16802

a) Electronic mail: shalaev@purdue.edu

**Abstract:**

We demonstrate a thermally tunable optical metamaterial with negative permeability working in the visible range. By covering coupled metallic nanostrips with aligned nematic liquid crystals (NLCs), the magnetic response wavelength of the metamaterial is effectively tuned through control of the ambient temperature, changing the refractive index of LC via phase transitions. By increasing the ambient temperature from $20\,^{\circ}$C to $50\,^{\circ}$C, the magnetic response wavelength shifts from 650nm to 632nm. Numerical simulations confirm our tests and match the experimental observations well.

Optical metamaterials with negative permeability and permittivity, which do not exist in nature, have attracted much attention in the past several years because of their unique properties [1]. Several structures such as split-ring resonators (SRRs), S-type or $\Omega$-type structures and fishnets have been developed to demonstrate negative index materials (NIMs) at near-infrared and visible optical wavelengths [2-6]. Materials can be characterized by their electric permittivity ($\varepsilon = \varepsilon' + i\varepsilon''$) and magnetic permeability ($\mu = \mu' + i\mu''$). Optical metamaterials with negative permeability and permittivity require $\varepsilon' < 0$ and $\mu' < 0$ to occur simultaneously within a wavelength range. However, the wavelength band exhibiting negative permeability is typically very narrow, while the band of negative permittivity can be much wider. Therefore, the key issue in fabricating negative refractive index materials is to implement and control the negative permeability.

There are several methods to control the negative permeability. The magnetic resonance wavelength of metamaterials is determined by their geometry and constituent materials. As a result, it is often possible to design the metamaterial structure for a magnetic resonance at a given wavelength [7]. On the other hand, the magnetic response wavelength can also be adjusted by changing the properties of the substrate [8]. However, the above methods are static and passive. It would be of great benefit to be able to change the behavior of the magnetic response dynamically.

Liquid crystals (LCs) have a large, broadband, optical anisotropy and refractive indices that are extremely sensitive to temperature and external electromagnetic fields [9]. Therefore, LCs are a good candidate for the dynamic control of permeability in metamaterials. Recently, Khoo *et al.* reported a theoretical analysis on tunable metamaterials using core-shell nanoparticles randomly dispersed in a matrix of LCs [10]. Werner *et al*. also proposed a tunable metamaterial incorporating a superstrate and substrate of nematic liquid crystals on a conventional NIM [11]. Zhao *et al* and Zhang *et al* experimentally showed that a negative-permeability, uni-planar, SRR array infiltrated with LCs can be reversibly controlled by external electric and magnetic fields [12, 13].

Thus far, all of the experimental results on LC-metamaterial tunability have been shown in the centimeter (microwave) frequency range. To date, no experimental demonstrations have shown tunability for the permeability of metamaterials in the optical wavelength range, and in particular within the visible range.

In this letter, we report a tunable, negative-permeability metamaterial for the visible light range. Control of the resonant wavelength is based on achieving a phase transition of nematic LCs via changing ambient temperature. The magnetic response wavelength can be effectively changed from 650nm to 632nm as the nematic liquid crystal undergoes a phase transition from the ordered phase to the isotropic phase as the temperature is raised over the phase transition temperature $T_c$ [~ 35°C for 5CB]. The resonant wavelength change is up to 3%, which is the best result in wavelength tuning of metamaterials observed to date.

It is already known that magnetic metamaterials (also called metamagnetics) consisting of arrays of paired thin silver strips exhibit a large magnetic response in the optical wavelength regime [7]. In our work, we used a similar structure to obtain an initial sample with a significant magnetic response. Electron-beam lithography and lift-off techniques were used to fabricate the metamagnetics samples on a glass substrate using the fabrication process as in Ref. [7]. Figure 1(a) shows the cross-sectional schematic of the metamagnetics samples. The figure shows a pair of thin silver strips (thickness 35nm) separated by an alumina spacer (thickness 60nm). The whole sandwich stack is trapezoidally shaped due to fabrication limitations. The structure is periodic with a periodicity of 300nm. Figure 1(b) and (c) show the field-emission scanning electron microscope (FE-SEM) image and the atomic force microscope (AFM) image of the sample. Note that two thin 10-nm alumina layers were deposited at the top and bottom of the $Ag-Al_2O_3-Ag$ sandwich stacks for fabrication stability.

The sample was optically characterized via far-field transmittance and reflectance spectral measurements using normally incident light at the primary polarization (i.e., the incident magnetic field is polarized along the set of strips, see Fig. 1(b)). The details of the experimental setup are described in Ref. [7]. The resulting spectra are shown as solid lines in Fig. 2(a), which clearly shows that there is a resonance dip in the transmission curve and a peak in the reflection curve near 595nm. This is the magnetic resonance resulting from an anti-symmetric current flow in the upper and lower strips, which forms a circular current and gives rise to a magnetic dipole response. This resonance induces the local minimum in the transmission spectrum and the local maximum in reflection. The absorption spectrum in Fig. 2(a) shows enhanced absorption near the resonance wavelength.

In addition to experimental characterization, we also investigated the properties of the metamagnetics sample by numerical simulations with a commercial finite-element package (COMSOL Multiphysics). The material properties of silver were taken from well-known experimental data [14], with the inclusion of an adjustable, wavelength-dependent loss factor [16]. The simulated results are shown as dashed lines in Fig. 2(a). We find that the theoretical and experimental results are consistent with each other. Using our simulation results, a negative effective permeability has been numerically retrieved. The value of $\mu'$ is about -1.5 at 595nm, which indicates that indeed the initial sample has a significant magnetic resonance, and large negative permeability is obtained. To illustrate the nature of the magnetic and electric resonances, we also simulated the field distribution at the resonance wavelength of a representative coupled nanostrip. The results are shown in Fig. 2(b), where the arrows represent electric displacement and the color map represents magnetic field. At the magnetic resonance wavelength, we find that the electric displacement forms a loop and produces an artificial magnetic moment, which results in negative permeability [7]. We also note a strong magnetic field inside the loop.

In order to investigate the tunability of the magnetic resonance, a layer of liquid crystal (5CB) was coated on top of the nanostrips at a temperature of 50$^\circ$C. The thickness of 5CB was about 400nm. Then, a glass substrate, which was initially coated with a 300-nm layer of robust PMMA, was sandwiched on top of the sample. The PMMA layer was used to align the extraordinary axis of the liquid crystal molecules parallel to the surface of the substrate and perpendicular to the direction of metamagnetics [15]. After forming the sandwiched structure, the whole sample was cooled down to room temperature (about 20°C). This sample (named Sample A) was then optically characterized by obtaining its transmittance spectrum using normally incident light at the primary linear polarization. The results of this characterization are shown in Fig. 3. When the sample was measured around 20°C, a transmission dip was observed at around 650nm. Compared with the same sample without LC (Fig. 2 (a)), a distinct red-shift is observed due to the increased effective permittivity of the dielectric environment. When the temperature was increased to 50°C, the transmission dip blue-shifted to 632nm. This change in resonance wavelength is completely reversible and repeatable. The shift in resonance wavelength is caused by a phase transition in the liquid crystal, which occurs at the ambient temperatures above 35°C. The LC is in the nematic phase at room temperature [12]. In the case of planar alignment by PMMA, the direction of the 5CB molecules is parallel to the surface of the glass substrate and perpendicular to the direction of the magnetic strips, with some thermal fluctuations away from the original alignment. So, the effective permittivity of 5CB for the primary polarization is close to $n_e = 1.73$, with some differences due to the thermal fluctuations. When the temperature is increased beyond 35°C, the 5CB liquid crystal molecules will transition from the nematic phase to the isotropic state. In the isotropic state, the LC molecules are randomly oriented and the material exhibits an isotropic refractive index that can be expressed as $n_{iso}^2 = \frac{2}{3}n_o^2 + \frac{1}{2}n_e^2 = 2.5$. The difference in refractive index between the two states is around $\Delta n = 0.15$. Normally incident light at the primary polarization experiences this change in refractive index during the phase transition. The decrease of the effective permittivity for the field at the primary polarization results in a blue-shift of the magnetic resonance [17].

We also simulated the dependence of the magnetic resonance wavelength on the LC phase transition using the structural parameters of the original sample and the COMSOL Multiphysics package. The results of these simulations are shown in Fig. 3 By matching the simulation results to experimental spectra, we found the effective refractive indices for LCs were 1.55 (nematic phase) and 1.4 (isotropic phase). Using these effective refractive indices, the resonance wavelength shift occurring as a result of the refractive index change $\Delta n = 0.15$ in the LC layer was also observed in simulation. The simulation results match the experimental observations very well. The effective refractive indices of 5CB used in our simulation for both the nematic and isotropic phase are smaller than those of real a LC medium. This is due to the air gap between the nanostrip surface and the LC medium. Because the nanostrips have narrow and deep gaps between two adjacent strips, it is hard for 5CB molecules to enter and fully fill the gap. Therefore, there will be a layer of air inside the gaps. This means that the effective refractive index of the dielectric environment for the nanostrips should be somewhat smaller than that of the pure LC. We must note here that the LC layer is assumed to be perfectly uniform in the numerical simulation, which is not quite the case for the experimental sample because the LC molecules can form in different droplets [18]. The high scattering between the different droplets in the LC layer results in the observed difference in the resonance strengths between the simulation and experimental results.

Another reference sample was also fabricated to verify the effect of the LC layer on the nanostrips. This sample was fabricated under the same conditions and using the geometries as the previous sample (Sample A). Instead of sandwiching a layer of LC in the structure, however, another layer of 10nm $Al_2O_3$ was coated on top of the nanostrips. The spectral transmission measurements (data not shown) indicate that there is no detectable shift of the resonance as the ambient temperature is increased from room temperature up to 50°C. Thus the temperature-tunable behavior of the LC-metamaterial sample is indeed caused by the inclusion of the LC molecules.

In summary, we have demonstrated a temperature-controlled, tunable negative permeability metamaterial working in the visible light regime by using a metamagnetics sample coated with a layer of liquid crystal molecules. Our results show that the magnetic resonance can be controlled reversibly by adjusting the environmental temperature. The resonance wavelength shift is caused by the temperature-induced phase transition of the LC material. As the phase transition of LCs can affect the refractive index over the whole optical wavelength spectrum and even into the microwave range, it is therefore possible to tune the magnetic response of metamaterials through the whole optical range. The dependence of the magnetic resonance wavelength on the phase of the LC layer is confirmed by simulations employing commercial finite-element software. The simulation matches the experimental results well.

**Acknowledgements**

This work was supported in part by ARO-MURI award 50342-PH-MUR. Useful discussions with Prof. Lavrentovich are appreciated.

**Figure 1**

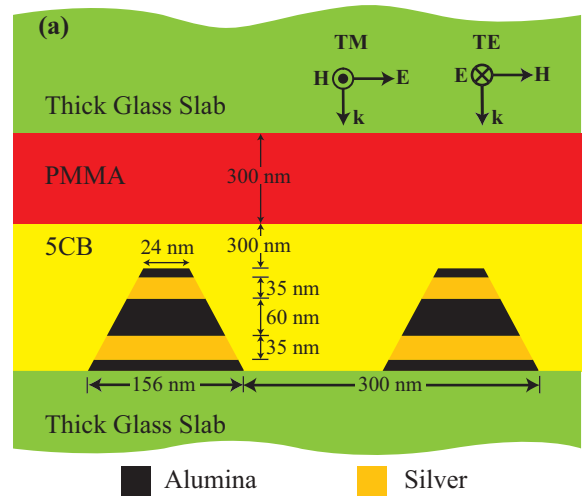

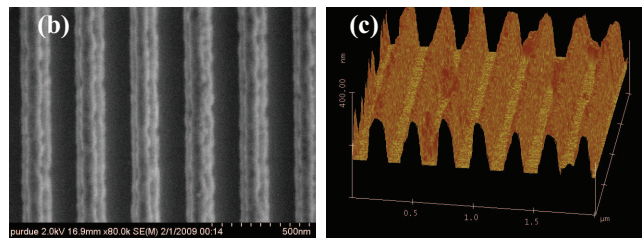

Fig. 1. Structure of the coupled nanostrip sample. (a) The cross-sectional schematic of arrays of coupled nanostrips; (b) The FE-SEM image of a typical sample; (c) The AFM image of the sample.

**Figure 2**

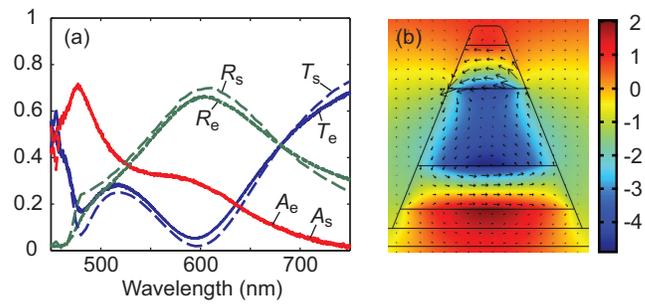

Fig. 2. (a)Transmission (T), reflection (R) and absorption (A), (including diffractive scattering) spectra under TM polarization for a typical coupled nanostrip sample with three characteristic wavelengths marked. Solid lines ($T_e$, $R_e$, and $A_e$) show the experimental data, and dashed lines ($T_s$, $R_s$, and $A_s$) represent simulated results. (b) Simulated electric displacement and magnetic field distributions at the magnetic resonance wavelength.

**Figure 3**

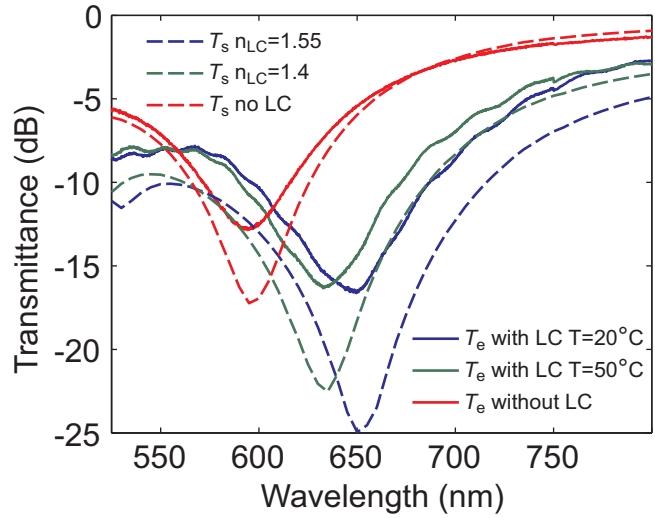

Fig. 3 Demonstration of a thermally tunable magnetic response in a metamaterial. Solid lines show the experimental data, and dashed lines represent simulated results without LCs (blue lines), with LCs on top at 20°C (black lines) and at 50 $^oC$ (red lines).